\documentclass[a4paper,fleqn,usenatbib]{mnras}

\usepackage{newtxtext,newtxmath}

\usepackage[T1]{fontenc}
\usepackage{ae,aecompl}

\usepackage{graphicx}	
\usepackage{amsmath}	
\usepackage{amssymb}	

\newcommand{\kms}{km~s$^{-1}$}

\newcommand{\cmn}{cm$^{-3}$}
\newcommand{\msun}{M$_{\odot}$}

\newcommand{\mum}{$\mu$m}
\newcommand{\lam}{$\lambda$}

\newcommand{\lya}{\mbox{Ly$\alpha$}}

\newcommand{\hi}{\mbox{H\,{\sc i}}}

\newcommand{\civ}{\mbox{C\,{\sc iv}}}
\newcommand{\ciii}{\mbox{C\,{\sc iii}}}
\newcommand{\cii}{\mbox{C\,{\sc ii}}}

\newcommand{\sixii}{\mbox{Si\,{\sc xii}}}
\newcommand{\siiv}{\mbox{Si\,{\sc iv}}}
\newcommand{\siiii}{\mbox{Si\,{\sc iii}}}

\newcommand{\siv}{\mbox{S\,{\sc iv}}}

\newcommand{\nv}{\mbox{N\,{\sc v}}}

\newcommand{\oviii}{\mbox{O\,{\sc viii}}}
\newcommand{\ovii}{\mbox{O\,{\sc vii}}}
\newcommand{\ovi}{\mbox{O\,{\sc vi}}}

\newcommand{\oi}{\mbox{O\,{\sc i}}}

\newcommand{\feii}{\mbox{Fe\,{\sc ii}}}

\newcommand{\fexxvi}{\mbox{Fe\,{\sc xxvi}}}
\newcommand{\fexxv}{\mbox{Fe\,{\sc xxv}}}

\newcommand{\mgx}{\mbox{Mg\,{\sc x}}}

\newcommand{\pv}{\mbox{P\,{\sc v}}}
\newcommand{\neviii}{\mbox{Ne\,{\sc viii}}}

\title[Does PDS 456 have a UV outflow at 0.3c?]{Does the X-ray outflow quasar PDS~456 have a UV outflow at 0.3c?}

\author[F. Hamann et al.]{
Fred Hamann$^{1},$\thanks{E-mail: fhamann@ucr.edu}
George Chartas$^{2}$,
James Reeves$^{3}$
and Emanuele Nardini$^{4}$
\\
$^{1}$ Department of Physics \& Astronomy, University of California, Riverside, CA 92507, USA \\
$^{2}$Department of Physics \& Astronomy, College of Charleston, Charleston, SC 29424, USA\\
$^3$ Astrophysics Group, School of Physical and Geographical Sciences, Keele University, Keele, Staffordshire, ST5 5BG, UK\\
$^4$ INAF -- Osservatorio Astrofisico di Arcetri, Largo Enrico Fermi 5, I-50125 Firenze, Italy\\
}

\date{Accepted XXX. Received YYY; in original form ZZZ}

\pubyear{2015}

\begin{document}
\label{firstpage}
\pagerange{\pageref{firstpage}--\pageref{lastpage}}
\maketitle

\begin{abstract}
The quasar PDS~456 (at redshift $\sim$0.184) has a prototype ultra-fast outflow (UFO) measured in X-rays. This outflow is highly ionized with relativistic speeds, large total column densities $\log{N_{\rm H}({\rm cm}^{-2})}>23$, and large kinetic energies that could be important for feedback to the host galaxy. A UV spectrum of PDS~456 obtained with the Hubble Space Telescope in 2000 contains one well-measured broad absorption line (BAL) at $\sim$1346~\AA\ (observed) that might be \lya\ at v$\;\approx{0.06c}$ or \nv\;\lam{1240} at v$\;\approx{0.08c}$. However, we use photoionisation models and comparisons to other outflow quasars to show that these BAL identifications are problematic because other lines that should accompany them are not detected. We argue that the UV BAL is probably \civ\;\lam{1549} at v$\;\approx{0.30c}$. This would be the fastest UV outflow ever reported, but its speed is similar to the X-ray outflow and its appearance overall is similar to relativistic UV BALs observed in other quasars. The \civ\ BAL identification is also supported indirectly by the tentative detection of another broad \civ\ line at v$\;\approx{0.19}c$. The high speeds suggest that the UV outflow originates with the X-ray UFO crudely 20 to 30~$r_g$ from the central black hole. We speculate that the \civ\ BAL might form in dense clumps embedded in the X-ray UFO, requiring density enhancements of only $\gtrsim$0.4 dex compared clumpy structures already inferred for the soft X-ray absorber in PDS~456.  The \civ\ BAL might therefore be the first detection of low-ionisation clumps proposed previously to boost the opacities in UFOs for radiative driving.
\end{abstract}




\section{Introduction}

Accretion disk outflows are an important part of the quasar phenomenon that might drive ``feedback" to regulate black hole growth and host galaxy evolution \citep[e.g.,][]{diMatteo05, Hopkins08, Hopkins10, DeBuhr12, Rupke13}. The outflows are often studied in the rest-frame UV via blueshifted broad absorption lines (BALs) or their narrower cousins called ``mini-BALs'' \citep[with a nominal boundary near full width at half minimum FWHM~$\sim 2000$ \kms ,][and refs. therein]{Weymann91, Korista93, Crenshaw03, Hamann04, Trump06, Knigge08, Gibson09}. These features appear most often at moderate velocity shifts v$\; < 0.1c$, but relativistic BALs and mini-BALs at v$\;\sim 0.1$--0.2$c$ have been measured in a small but growing number of quasars \citep{Hamann97b, Hamann13, Paola08, Paola11, Rogerson16}. 

The basic physical properties for these outflows can be difficult to determine due to limited wavelength coverage and line saturation that is masked by partial line-of-sight covering of the background light source(s). The strongest measured UV lines\footnote{Throughout this paper we treat unresolved doublets as single lines, such that, for example, \civ\ \lam 1548,1551 becomes \civ \lam 1549 with a summed oscillator strength (for column density estimates).} are typically \civ\ \lam 1549, \nv\ \lam 1240, and \ovi\ 1034, indicating moderate to high degrees of  ionisation. The maximum ionisations are not known, but there might be a wide range. \ovi\ absorption tends to be as strong or stronger than \civ\ \citep[Section 3.4 below, also][Herbst et al., in prep.]{Baskin13,Moravec17} and higher-ionisation lines such as \neviii\ \lam 774 and \mgx\ \lam 615 have been observed in a few cases with suitable spectral coverage \citep[e.g.,][]{Hamann97,Telfer98, Arav01}. Measurements of the low-abundance line \pv\ \lam 1121 help to overcome the saturation issues to reveal generally large total hydrogen column densities, $\log N_{\rm H}({\rm cm}^{-2}) \gtrsim 22.3$, across a wide range observed BAL strengths \citep[][but see also Arav et al. 2001]{Hamann98, Leighly09, Leighly11, Borguet12, Capellupo14, Capellupo17, Moravec17}.  

X-ray observations have revealed another variety of ultra-fast outflows (UFOs) that reach relativistic speeds in luminous quasars \citep[][and refs. therein]{Chartas02, Chartas09, Reeves09, Gofford13, Gofford15, Tombesi10, Tombesi13}. UFOs are also challenging to study because they are highly variable and highly-ionised to the point where the only strong absorption features appear at X-ray wavelengths. They are often characterized by Fe K-shell absorption features with derived total column densities in the range $\log N_{\rm H}({\rm cm}^{-2}) \sim 22$ to 24 (see refs. above). UFOs also appear to have generally very large kinetic energies, sufficient to drive feedback effects in the quasar host galaxies \citep{Tombesi12, Tombesi13, Gofford14, Gofford15, Reeves14}. 

The quasar PDS~456 (at redshift $z_e\approx 0.184$) has the best-studied example of a powerful, relativistic X-ray UFO \citep[e.g.,][]{Reeves03, Reeves09, Reeves14, Reeves16, Nardini15, Gofford14,Matzeu17}. It is the most luminous quasar in the local universe, with bolometric luminosity $L\sim 10^{47}$ ergs s$^{-1}$ and estimated black hole mass $\sim (1-2)\times 10^9$ \msun\ that together indicate an accretion rate relative to Eddington that is $L/L_{Edd}\gtrsim 0.3$ and perhaps near unity \citep{Nardini15}. The X-ray absorber is complex and highly variable, with two main components. The main component measured via Fe K-shell absorption has speeds in the range v$\;\sim 0.25$--0.34$c$, very high degrees of ionisation featuring \fexxv\ and \fexxvi , and large total column densities $\log N_{\rm H}({\rm cm}^{-2}) \gtrsim 23$ \citep{Reeves14, Gofford14, Gofford15, Nardini15}. Its radial distance from the black hole is estimated at a few hundred gravitational radii \citep[based on absorber variability, e.g.,][]{Nardini15}. The kinetic power of this outflow is also remarkably large, crudely $\sim$20 percent of the bolometric luminosity, which is well above the threshold needed for feedback to the host galaxy \citep[][and refs. therein]{Gofford14, Nardini15}. The second outflow component measured in soft X-rays has somewhat lower speeds, v$\;\sim 0.17$--0.27$c$, lower ionisations, and column densities in the range $\log N_{\rm H}({\rm cm}^{-2}) \sim 22$ to 23 \citep{Reeves16}. There is evidence for time-variable covering fractions in this absorber that might be indicative of small dense clumps or substructures embedded in the overall X-ray outflow \citep{Matzeu16}. 

PDS~456 also has signatures of outflow in the UV. A UV spectrum obtained in 2000 by \cite{OBrien05} using the Space Telescope Imaging Spectrometer (STIS) on board the Hubble Space Telescope (HST) revealed a highly blueshifted \civ\ broad emission line at speeds near v$\;\sim 5000$ \kms\ plus a single BAL that might plausibly be \lya\ or \nv\ at v$\sim 0.06c$ or $\sim$0.08$c$, respectively. These UV features provide further evidence for exotic mass loss from PDS~456 across a wide range of spatial scales, from the X-ray UFO that is believed to originate near the black hole at radii of $\sim20$--$30\, r_g$ \citep[gravitational radii, corresponding to $\sim$0.001--0.0015 pc][]{Nardini15, Matzeu17} to an outflow-dominated \civ\ broad emission-line region that we place crudely at $\sim$0.3 pc (based on scaling relations with luminosity, \citealt{Kaspi05}, for a 10$^9$ \msun\ black hole). The UV BAL is an important component to this outflow picture, but its measured speed and physical nature depend critically on the line identification. 

In this paper, we reexamine the UV BAL in PDS~456 with the main result that it is likely to be \civ\ \lam 1549 at v$\;\approx 0.30c$. This would be the fastest UV outflow line ever reported but similar in speed to the X-ray UFO in this quasar. Throughout this paper, we adopt a redshift for PDS~456 of $z_e = 0.18375$ based on the emission line [\feii ] 1.6435 \mum\ measured by \cite{Simpson99}. We describe two archival HST spectra of PDS~456 in Section 2 below. Section 3 presents our analysis and  measurements of the observed BAL, comparisons to photoionisation models, comparisons to UV BALs in other quasars, and a discussion of the plausible BAL identifications. Section 4 presents a summary and discussion of the results. 

\section{HST Spectra}

Figure 1 shows the spectrum of PDS~456 obtained by \cite{OBrien05} in 2000 using HST-STIS with the G140L and G230L gratings. We obtained this spectrum and another HST spectrum measured in 2014 (described below) from the Mikulski Archive for Space Telescopes (MAST). We present them here without further processing. The STIS G140L grating provided wavelength coverage from 1137 \AA\ to 1715 \AA\ at resolutions ranging from $\sim$310 \kms\ to $\sim$200 \kms , while STIS G230L covered 1580 \AA\ to 3148 \AA\ at resolutions from $\sim$640 \kms\ to $\sim$300 \kms\ (see \citealt{OBrien05} for more details). The spectrum plotted in Figure 1 is combined from the two gratings by masking out the extreme ends of the wavelength coverage to avoid excessive noise and then calculating variance-weighted average fluxes at the remaining wavelengths of overlap. 

\begin{figure}
	\includegraphics[width=\columnwidth]{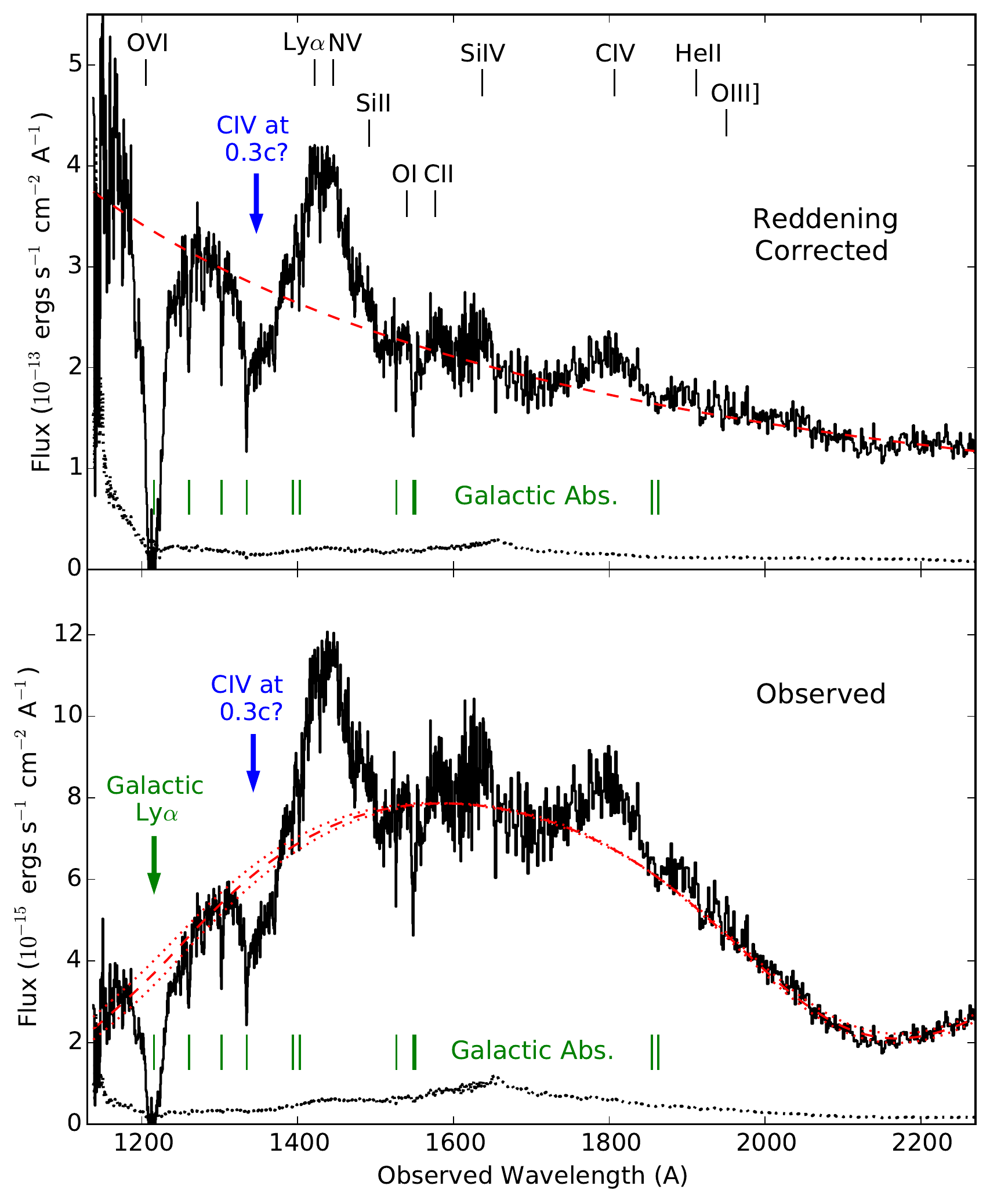}
\vskip -5pt
    \caption{HST STIS spectrum of PDS 456 from 2000 plotted as observed (bottom panel) and reddening-corrected using a Galactic extinction curve with $E(B-V) = 0.45$ (top panel). The grey dotted curves are the corresponding error spectra. The red dashed curves are an approximate fit to the continuum using a power law with index $\alpha_{\lambda} = -1.68$ shown with (bottom panel) and without (top) Galactic reddening. The red dotted curves in the bottom panel show the same power law reddened alternatively by $E(B-V) = 0.42$ and 0.48. The BAL we attribute to \civ\ at v$\;\approx 0.30c$ is labeled above the spectrum at 1346 \AA . Some broad emission-line wavelengths are marked across the top. Galactic absorption lines (including strong damped \lya\ at 1216 \AA ) are marked by green vertical lines near the bottom. See Section 2.}
    \label{fig:example_figure}
\end{figure}

The UV spectrum of PDS~456 is substantially reddened by dust in our Galaxy (due to the quasar's sky position near the Galactic plane). Previous studies based on visible spectra and photometry indicate that the visual extinction towards PDS~456 is $A_V\sim 1.5$ magnitudes, corresponding to selective extinction $E(B-V) \sim 0.48$ for a standard Galactic reddening curve \citep{Torres97,Simpson99}. We obtain an independent estimate of the reddening by fitting the UV continuum with a single power law modified by the Galactic extinction curve from \cite{Cardelli89} with $R_V = 3.1$. The fit is constrained by the median flux in wavelength windows between $\sim$1170 \AA\ and $\sim$2400 \AA\ that avoid strong emission and absorption lines. The dashed red curve in Figure 1 shows our preferred fit using $E(B-V) = 0.45$ and power law index $\alpha_{\lambda} = -1.68$ (for $F_{\lambda}\propto \lambda^{\alpha_{\lambda}}$). An important feature of Galactic extinction at these wavelengths is the ``bump'' at $\sim$2175 \AA . We experimented with different fit parameters (e.g., the dotted red curves in Figure 1), but the values of $E(B-V)$ and the power law slope are well constrained\footnote{We note, however, that the observed spectrum is poorly fit at wavelengths $\gtrsim$2400 \AA\ (observed). This might be caused by reddening in the quasar environment, a break in the intrinsic/emitted spectral slope, and/or blended \feii\ emission lines that are known to be strong in PDS~456 \citep{Simpson99}. We do not investigate this further because our goal is simply to define a continuum for analysis of the outflow lines at shorter wavelengths (Section 3).} by the data because Galactic reddening curves with a strong 2175 \AA\ bump suppress the flux at both ends of the spectral coverage shown in Figure 1. The top panel in this figure shows the reddening-corrected spectrum of PDS~456 together with the best-fit power law.

Figure 2 shows a previously-unpublished spectrum of PDS~456 obtained in 2014 with HST-COS using the G140L grating (PI: O'Brien). This spectrum (shown by the red curve) is plotted on top of the HST-STIS spectrum (black curve) from Figure 1 with no reddening corrections. It is combined from separate exposures totaling 5186 seconds. The spectral resolution ranges from roughly 200 \kms\ to 100 \kms\ from blue to red across the wavelengths shown in Figure 2. A gap in the COS wavelength coverage from $\sim$1170 \AA\ to $\sim$1265 \AA\ avoids Galactic/geocoronal \lya\ absorption/emission. The blue arrows in Figures 1 and 2 mark the positions of observed or expected absorption lines discussed in Section 3 below 

\begin{figure}
	\includegraphics[width=\columnwidth]{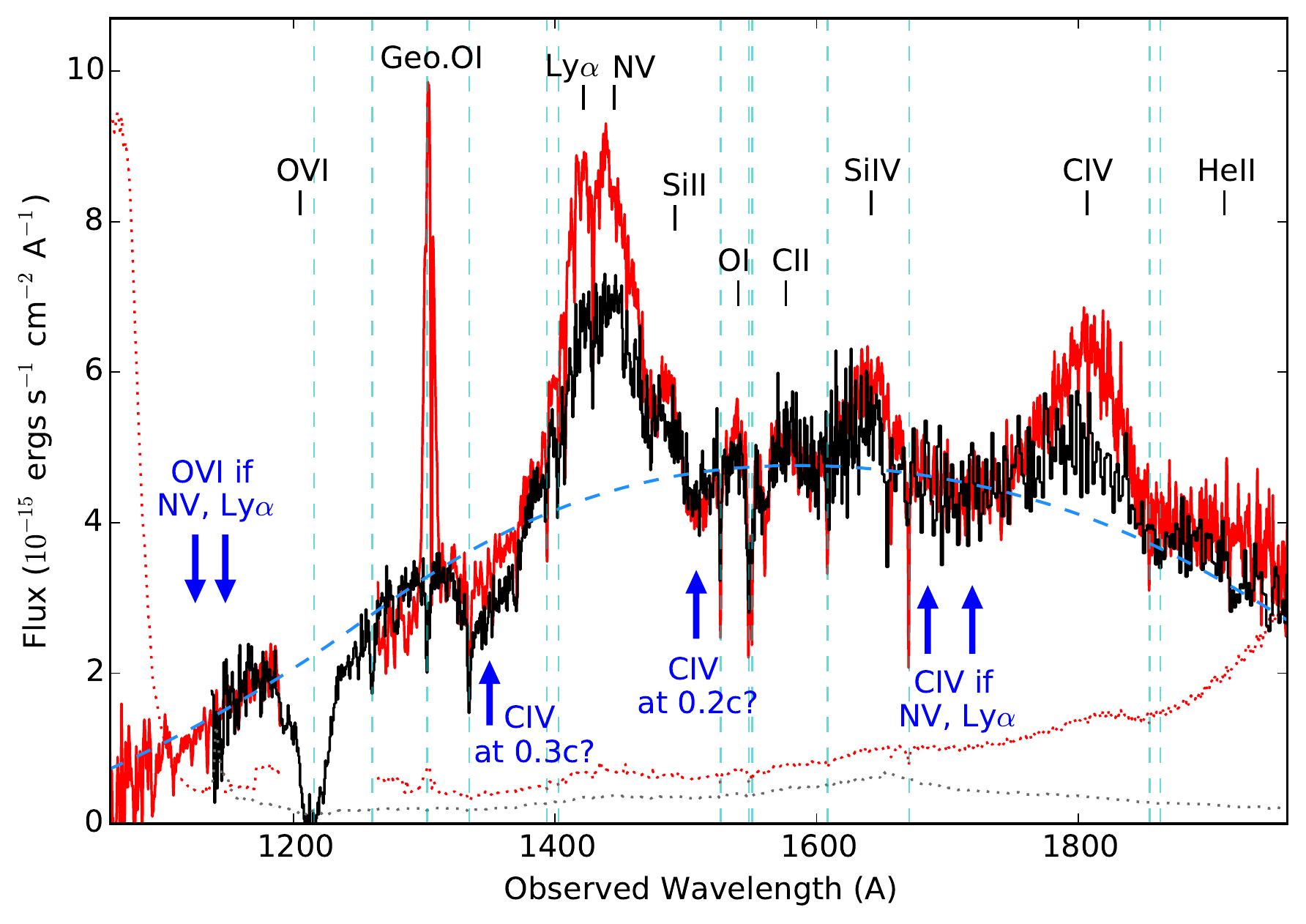}
\vskip -5pt
    \caption{HST spectra of PDS~456 obtained with STIS in 2000 (black curve) and COS in 2014 (red) plotted at observed wavelengths, uncorrected for reddening. The red and grey dotted curves are the corresponding error spectra (1$\sigma$ uncertainties per pixel). The STIS spectrum is scaled vertically by a factor $\sim$0.60 to match the COS spectrum approximately in the continuum. The dashed blue curve shows our fit to the STIS spectrum from Figure 1. The BAL we attribute to \civ\ \lam 1549 at 0.30$c$ is labeled below the spectrum at $\sim$1346 \AA . Also marked are the expected positions of \ovi\ \lam 1034 and \civ\ BALs if the observed BAL is attributed, instead, to \nv\ \lam 1240 at 0.08$c$ or \lya\ at 0.06$c$ (e.g., ``if \nv , \lya "). We tentatively identify another weak BAL at $\sim$1513 \AA\ as \civ\ at 0.19$c$. Dashed blue vertical lines mark the positions of Galactic absorption features. The narrow emission spike at 1304 \AA\ is geocoronal \oi . See Figure 1 and Section 2 for additional notes.}
    \label{fig:example_figure}
\end{figure}

\section{Analysis}

\subsection{The UV BAL at 1346 \AA}

The HST-STIS spectrum in Figure 1 clearly shows a BAL at $\sim$1346 \AA\ (observed), as reported by \cite{OBrien05}. Figure 3 plots this spectrum again after normalizing by the reddened power law in Figures 1 and 2. We fit the BAL in this normalized spectrum using a simple Gaussian optical depth profile of the form
\begin{equation}
\tau_{\rm v} \ = \ \tau_o\, e^{-{\rm v}^2/b^2}
\end{equation}
where $\tau_o$ is the line-center optical depth, v is the velocity shift from line center, and $b$ is the doppler parameter that sets the line width. The BAL fit shown by the magenta curve in Figure 3 yields $\tau_o = 0.35\pm 0.01$, $b = 5135\pm 203$ \kms , and observed line-center wavelength $\lambda_{o} = 1345.7\pm 0.6$ \AA\ (where the errors quoted are 1$\sigma$ uncertainties returned by the line-fitting routine). The full width at half minimum of this fitted profile is FWHM$\;\approx 8550\pm 340$ \kms . In the quasar frame, the fit has rest equivalent width REW$\; = 10.8\pm 0.6$ \AA . 

\begin{figure}
\includegraphics[width=\columnwidth]{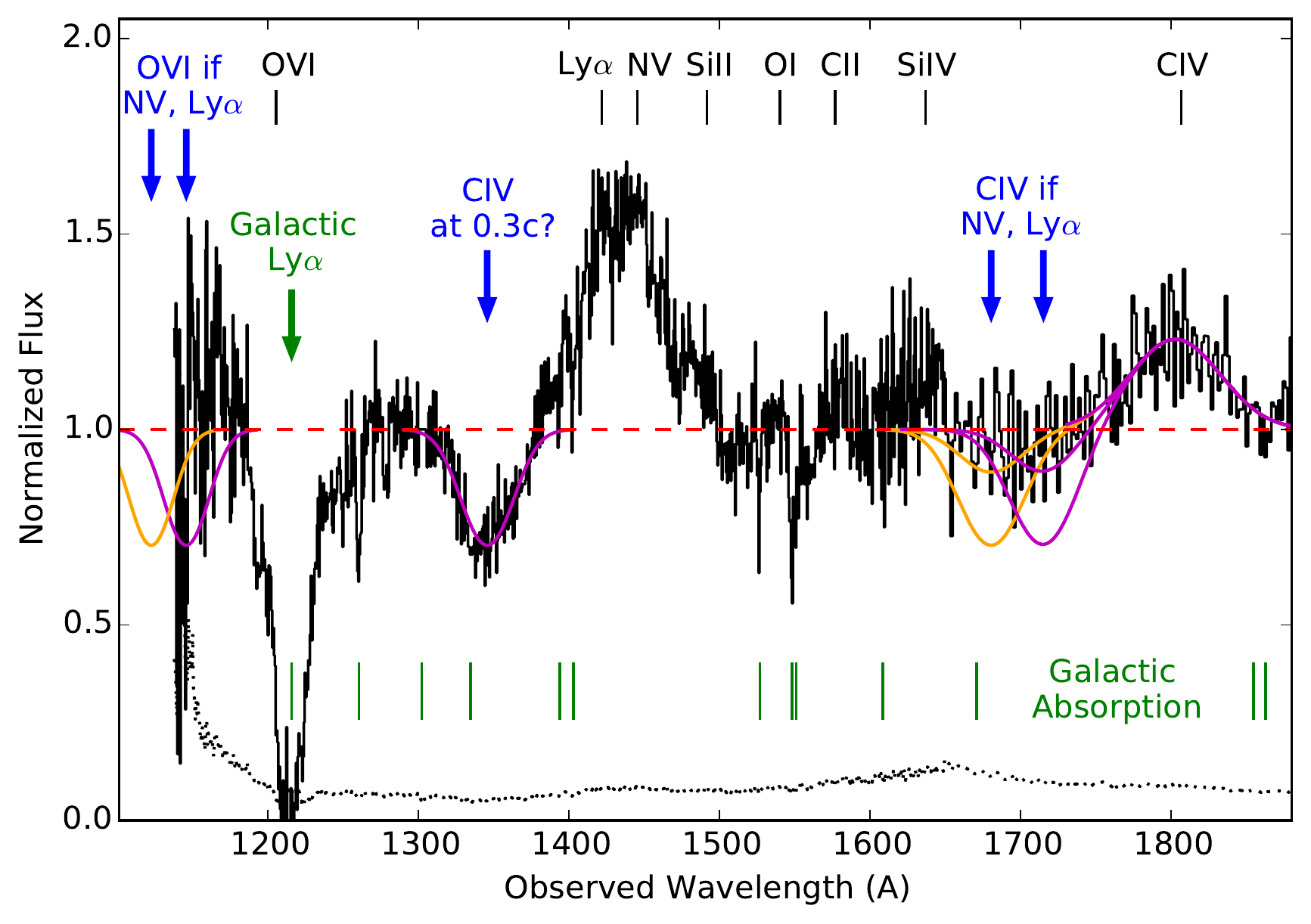}
\vskip -5pt
  \caption{HST-STIS spectrum of PDS 456 in the observed frame, normalized by the continuum fit shown in Figure 1. The smooth magenta and orange curves show our fit to the observed BAL at 1346 \AA\ and then that fit transposed to the wavelengths of \ovi\ and \civ\ assuming the measured BAL is \lya\ (magenta) or \nv\ (orange). The weaker profiles drawn at the \civ\ positions show the transposed fit at 1/3 its measured strength to illustrate approximate upper limits to the actual absorption there. The absence of absorption at the transposed fit positions argues against the \lya\ and \nv\ identifications for the observed BAL (Section 3.5 and Section 3.6). Other labels are the same as Figures 1 and 2. Narrow Galactic absorption lines of \oi\ \lam 1304 and \cii\ \lam 1335 have been removed by interpolation for clarity.}
    \label{fig:example_figure}
\end{figure}

Our study concerns identification of the $\sim$1346 \AA\ BAL trough. One possibility favored by \cite{OBrien05} is \lya\ 1216 at velocity shift v$\;\approx 0.06c$ (18,000 \kms ). Another is \nv\ \lam 1240 at v$\;\approx 0.08c$. These identifications seem reasonable because the observed BAL is on the blue side of the \lya --\nv\ emission-line blend and the inferred velocities are in a normal range for BAL outflows in other quasars. However, we will argue below (Section 3.5 and Section 3.6) that the \lya\ or \nv\ identifications are problematic because they are not accompanied by \civ\ and \ovi\ \lam 1034 BALs at the same velocity shifts. The absence of these other lines is illustrated by the other orange and magenta curves in Figure 3, which show our fit to the observed BAL transposed to the positions expected for accompanying \civ\ and \ovi\ BALs . At the \civ\ positions, the transposed BAL fit is shown twice -- once at full strength and once at 1/3 of its  measured strength at 1346 \AA . We estimate that the 1/3 scalings represent approximate upper limits to \civ\ BALs that might accompany a \nv\ or \lya\ BAL at 1346 \AA . We will argue below that the absence \ovi\ and \civ\ absorption at these wavelengths suggests that the BAL at $\sim$1346 \AA\ is actually \civ\ \lam 1549 at v$\;\approx 0.30c$. 

We test these line identifications using photoionisation models (Section 3.3) and comparisons to BALs/mini-BALs observed in other quasars (Section 3.4). First we estimate ionic column densities for each possible identification. These estimates follow from the fitted Gaussian optical depth profile by
\begin{equation}
N_i \ = \ {{m_e\, c}\over{\sqrt{\pi}\, e^2}}~{{b\,\tau_o }\over{f\lambda_r}}
\end{equation}
where $N_i$ is the ionic column density, $f$ and $\lambda_r$ are the oscillator strength and rest-frame wavelength of the transition, and we assume implicitly that the ground-state column density equals the ionic column density. Thus we find that, if the observed BAL is \lya , it corresponds to a neutral hydrogen column density of $\log N_{\rm HI} ({\rm cm}^{-2}) \approx 15.35$. If the BAL is \nv\ or \civ , then the column density is $\log N(\nv)  ({\rm cm}^{-2})  \approx 15.62$ or $\log N(\civ )  ({\rm cm}^{-2}) \approx 15.44$, respectively. The uncertainties in these estimates depend mainly on the continuum placement used for the BAL profile fit. Experiments with different continua suggest that the maximum uncertainties are roughly 0.1 dex. 

It is important to note that the line optical depth and column densities derived above are only lower limits because BAL outflows often exhibit saturation with partial line-of-sight covering of the background light source. This can lead to weak/shallow absorption-line troughs even if the optical depths are large \citep[e.g.,][and refs. therein]{Hamann98, Hamann04, Arav05, Capellupo17, Moravec17}. In one recent study, Herbst et al. (in prep.) used median composite spectra of BAL quasars from the Baryon Oscillation Sky Survey (BOSS, \citealt{Dawson13,Ross12}, part of the Sloan Digital Sky Survey-III, SDSS-III, \citealt{Eisenstein11}) to show that the low-abundance doublet \pv\ \lam 1118,1128 is often present and that widely-separated doublets like \pv , \siiv\ \lam 1393,1402, and \ovi\ \lam 1032,1038 have $\sim$1:1 doublet ratios across a wide range of BAL strengths. These results indicate that all of the commonly measured BALs of abundant ions in quasar spectra, e.g., \civ\ \lam 1549, \nv\ \lam 1240 and \ovi\ \lam 1034, are {\it typically} saturated and the observed depths of the BAL troughs are controlled mainly by the line-of-sight covering fractions. 

\subsection{Other UV Outflow Features}

In addition to the UV BAL at 1346 \AA , there are two other features in the 2000 HST-STIS spectrum that point to extreme outflows in PDS~456. First, as noted previously by \cite{OBrien05}, the broad \civ\ emission line is extremely blueshifted. Our simple Gaussian fit to this feature shown in Figure 3 indicates that the line centroid is blueshifted by $5200\pm 450$~\kms\ (consistent with the \citealt{OBrien05} measurement). This fit also yields a rest equivalent width of REW~$= 14.7\pm 1.8$ \AA\ and a large velocity width of FWHM~$=11770\pm 1060$ \kms\ (not corrected for the $\sim$500~\kms\ doublet separation in \civ ). The low-ionisation emission lines \oi\ \lam 1304 and \cii\ \lam 1335 are poorly measured in this spectrum compared to \civ , but they are clearly less blueshifted. We crudely estimate their centroid blueshifts by visual inspection to be $700\pm 200$~\kms .

The large \civ\ emission-line blueshift identifies an outflow-dominated broad emission-line region with an unusually high outflow speed. At $v\approx 5200\pm 450$~\kms , the \civ\ blueshift is outside of the range of values reported by \cite{Coatman16} for a sample of $\sim$31,157 luminous SDSS quasars \citep[see also][]{Richards11}. However, we can crudely estimate by extrapolation from their Figure 1 that the \civ\ blueshift in PDS~456 is in the upper $\sim$0.1 percent of luminous quasars. To our knowledge, it is matched in the literature only by a few luminous extremely red quasars (ERQs) discovered in SDSS-III/BOSS to have prodigious outflows include BALs in the UV and highly blueshifted [OIII] \lam 5007 emission lines \citep[][Perrotta et al., in prep.]{Hamann17,Zakamska16}. We will present a more thorough analysis of the \civ\ blueshift in a future paper. Here we note simply that large blueshifts are known to correlate with small emission-line REWs and intrinsically weak X-ray emissions \citep[compared to other quasars/AGN at similar luminosities, e.g.,][]{Leighly04b, Leighly07b, Wu11j, Wu12j, Richards11, Luo15}. In addition, the work by \cite{Coatman16} supports speculation in the ERQ studies that large blueshifts and other prominent outflow features are related to high accretion rates (high Eddington ratios) in the quasars. 

Another tentative outflow feature in the 2000 HST-STIS spectrum is a weak BAL marginally detected at $\sim$1513 \AA\ (observed, see Figures 2 and 3). The reality of this feature depends on the continuum placement, but it is clearly below our best guess at the continuum and might be stronger in the HST-COS 2014 spectrum (Figure 2). There is also tentative absorption near $\sim$1560 \AA , but this feature is blended with telluric \civ\ absorption at 1549 \AA\ and it appears to be much narrower than the BAL candidate at $\sim$1513 \AA . Fits to the BAL candidate at 1513 \AA\ are poorly constrained, but multiple trials indicate a central wavelength of $\lambda_o \approx 1513\pm 2$ \AA , FWHM$\;\approx 3800\pm 1000$ \kms , and REW$\; \approx 1.1\pm 0.5$ \AA\ in the quasar frame. If this BAL is real, the only plausible identification is \civ\ at v$\; \sim 0.19c$ (The leading alternative, \siv\ \lam 1398 at v$\;\sim 0.09c$ is ruled out by the absence of \civ\ absorption at the same velocity shift. See \citealt{Paola11,Hamann13}). 

\subsection{Cloudy Simulations}

We use the photoionisation and spectral synthesis code Cloudy version 17.00 \citep{Ferland13,Ferland17} to predict absorption-line strengths for different physical conditions that might produce the observed BAL at $\sim$1346 \AA . The calculations assume twice solar metallicity and a fixed hydrogen density of $n_H = 10^7$ \cmn . The specific density has no bearing on our results \citep[e.g.,][]{Hamann97d}. The metallicity can affect mainly the metal line strengths relative to the Lyman lines. We choose twice-solar metallicity to be crudely consistent with other studies of the outflows and broad emission-line environments of luminous quasars \citep[e.g.,]{Hamann99, Dietrich02, Warner04, Nagao06, Gabel05}. The clouds are irradiated by a standard quasar ionising spectrum consistent with measurements of PDS~456 \citep[e.g.,][]{Nardini15}. This spectrum is defined by power laws across optical-UV and X-ray wavelengths with slopes of $\alpha_{uv}=-0.5$ and $\alpha_x=-1.3$, respectively (for $f_{\nu}\propto\nu^{\alpha}$). The two power laws are joined smoothly in the far-UV by an exponential Wien function with temperature $T$ = 250,000 K. The relative strengths of the UV and X-ray spectral segments are set by a two-point power-law index between 2500 \AA\ and 2 keV equal to $\alpha_{ox} = -1.8$ \citep{Strateva05,Steffen06}. This spectrum is defined in Cloudy by the command: {\tt AGN T=250000K, a(ox)=-1.8, a(uv)=-0.5, a(x)=-1.3}. The flux of hydrogen ionising radiation incident on the model clouds is set by the ionisation parameter, 
\begin{equation}
U \ \equiv \ {{Q_H}\over{4\pi c R^2\, n_H}}
\end{equation}
where $R$ is the distance from the quasar light source and $Q_H$ is the total emitted luminosity of hydrogen-ionising photons (\#/s). Other aspects of the calculations are described below. 

\subsection{A Comparison Outflow Sample}

Here we construct a sample of outflow quasars from SDSS-III/BOSS that can be useful analogues to test the BAL identification in PDS~456. We specifically use data from the BOSS quasar catalog for data release 12 \citep[DR12,][]{Paris17} to select quasars with the following properties: 1) We require that the BOSS spectra have signal-to-noise ratios SNR $>$ 5 at 1700 \AA\ (quasar frame). 2) The emission-line redshifts must be $z_e > 2.4$ to place the outflow lines of \civ , \nv\ and \lya\ within the BOSS wavelength coverage at speeds up to at least $\sim$0.1$c$. Nearly 2/3 of the selected quasars have $z_e > 2.6$ such that \ovi\ \lam 1034 is also covered. 3) The quasars must have broad \civ\ outflow lines (BALs or mini-BALs) characterized by balnicity index BI $<$ 3000 \kms\ and absorption index AI $>$ 1000 \kms , both measured at $>$4$\sigma$ confidence \citep[see][and refs. therein for definitions of BI and AI]{Paris17}. These parameter limits are designed to exclude strong/deep \civ\ BALs (with large BI) as well as very narrow \civ\ lines with FWHMs $<$ several hundred \kms\ (with small AI) that do not resemble the observed feature in PDS~456. Note that broad outflow lines are recorded in the BOSS quasar catalog only at velocity shifts $v\,\lesssim 0.1c$. 4) The {\it minimum} velocity recorded on the red side of the BAL/mini-BAL troughs is v$\;>$~6000 \kms\ (as specified by the AI integration limit {\tt vmin\_civ\_450} $>$ 6000). This avoids broad lines near v$\;\sim 0$ that can have different profiles and other properties compared to higher-velocity BALs and mini-BALs. The specific value of 6000 \kms\ is a compromise that requires a substantial velocity shift while still maintaining a large sample size. Finally, 5) we reject a small fraction of the quasars ($\sim$10 percent) that have damped \lya\ absorption (DLAs) or strong sub-DLAs at the wavelengths of interest (based on our own visual inspections of the spectra). 

The final sample includes 641 quasars at median redshift $\left<z_e\right> = 2.84$ and median absolute $i$-band magnitude $\left<M_i\right> = -26.8$. We visually inspect all of the quasar spectra to assess the strengths of prominent outflow lines for comparison to PDS~456. At these redshifts and velocity shifts, the \nv , \lya , and \ovi\ outflow lines are inevitably blended with unrelated absorption features in the \lya\ forest. However, in most cases, the outflow lines are clearly discernible because they are much broader than the forest features. Figure 4 shows spectra of four quasars we find to have the strongest well-measured \nv\ outflow lines relative to \civ\ in our BOSS sample. These extreme cases with large \nv /\civ\ are useful for our discussions of the BAL identification in PDS~456 (e.g., Section 3.6). 
\begin{figure}
\includegraphics[width=\columnwidth]{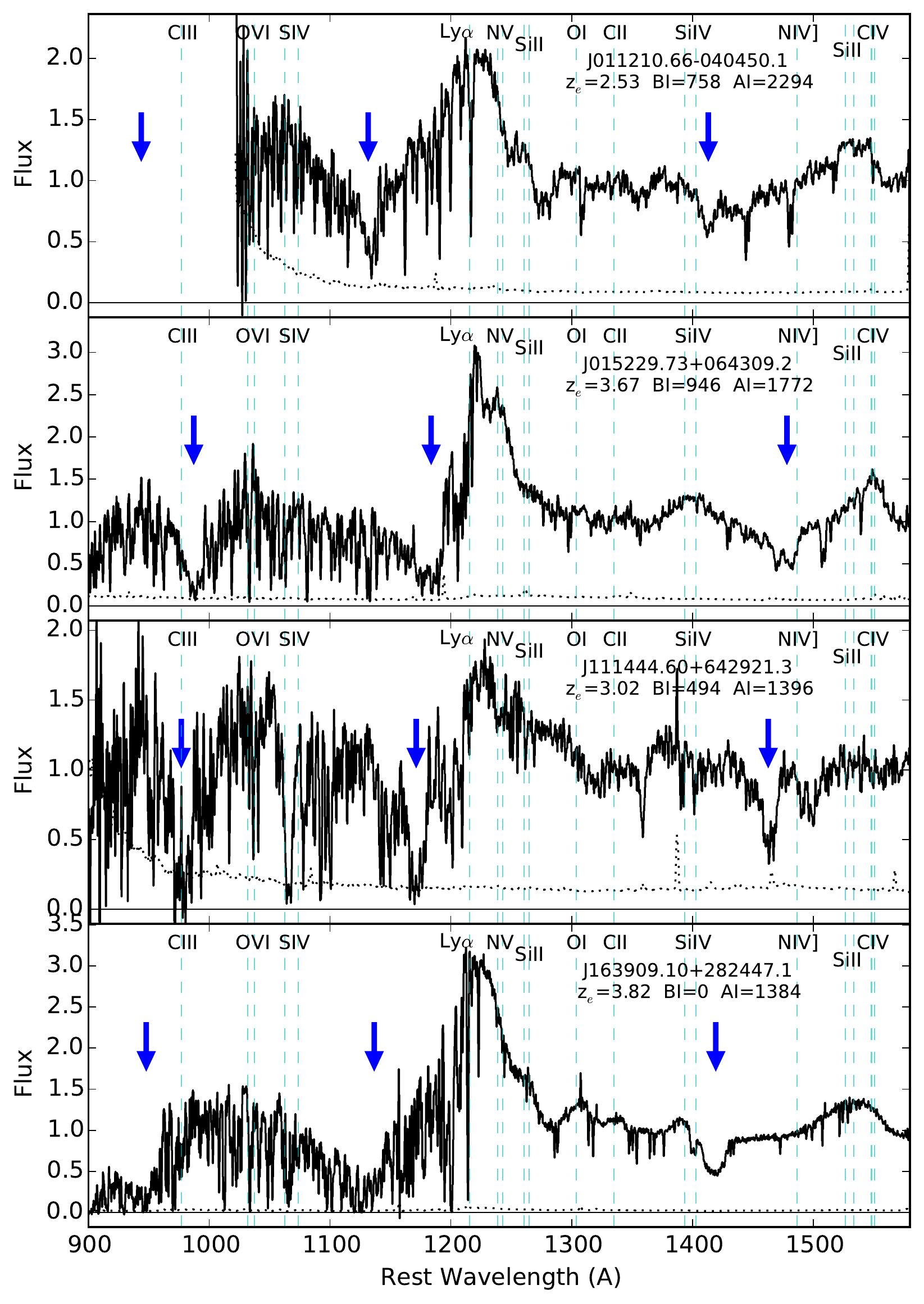}
\vskip -5pt
  \caption{Normalized spectra in the quasar frame of four outflow quasars from our BOSS sample with the strongest well-measured \nv\ absorption lines relative to \civ . Blue arrows mark the broad outflow absorption lines of \ovi , \nv , and \civ\ from left to right, respectively. \nv\ and \ovi\ suffer from varying amounts of contamination in the \lya\ forest (e.g., in the bottom panel there is Lyman limit absorption at wavelengths $\lesssim$912 \AA\ and the deepest part of the broad \nv\ trough is enhanced and offset from the absorption minima in \civ\ and \ovi\ due to \lya\ forest contamination). Prominent broad emission lines are labeled across the top. Also listed are the quasar names, emission-line redshifts, $z_e$, and absorption indices, BI and AI, from the BOSS DR12 quasar catalog. See Section 3.4}
    \label{fig:example_figure}
\end{figure}

It is also useful to consider the median outflow line properties in the BOSS sample. Thus we construct a median composite spectrum in the outflow absorber frame for all 641 quasars. This composite ``averages out'' the \lya\ forest contamination to reveal the typical outflow line strengths and profiles across the sample. The result is shown in Figure 5. We derive an absorber redshift for each quasar from the wavelength of minimum flux between the limits of the \civ\ AI integration (e.g., between {\tt vmin\_civ\_450} and {\tt vmax\_civ\_450}) in smoothed versions of the BOSS spectra. The composite is then constructed by normalizing each spectrum to a continuum flux near 1700 \AA\ and then shifting to the absorber frame before calculating the median. We remove broad emission lines from the final outflow quasar composite spectrum by dividing by another composite spectrum of $\sim$7700 BOSS quasars with similar redshifts and absolute magnitudes ({\tt Mi}) but {\it without} broad outflow lines in their spectra. The redshifts of these non-outflow quasars are randomly displaced to match the distribution of shifts used to place the outflow quasars in the absorber frame. A simple division then yields the final composite spectrum shown in Figure 5. See \cite{Baskin13}, \cite{Baskin15}, and Herbst et al. (in prep.) for more discussion of this procedure.  

\begin{figure}
\includegraphics[width=\columnwidth]{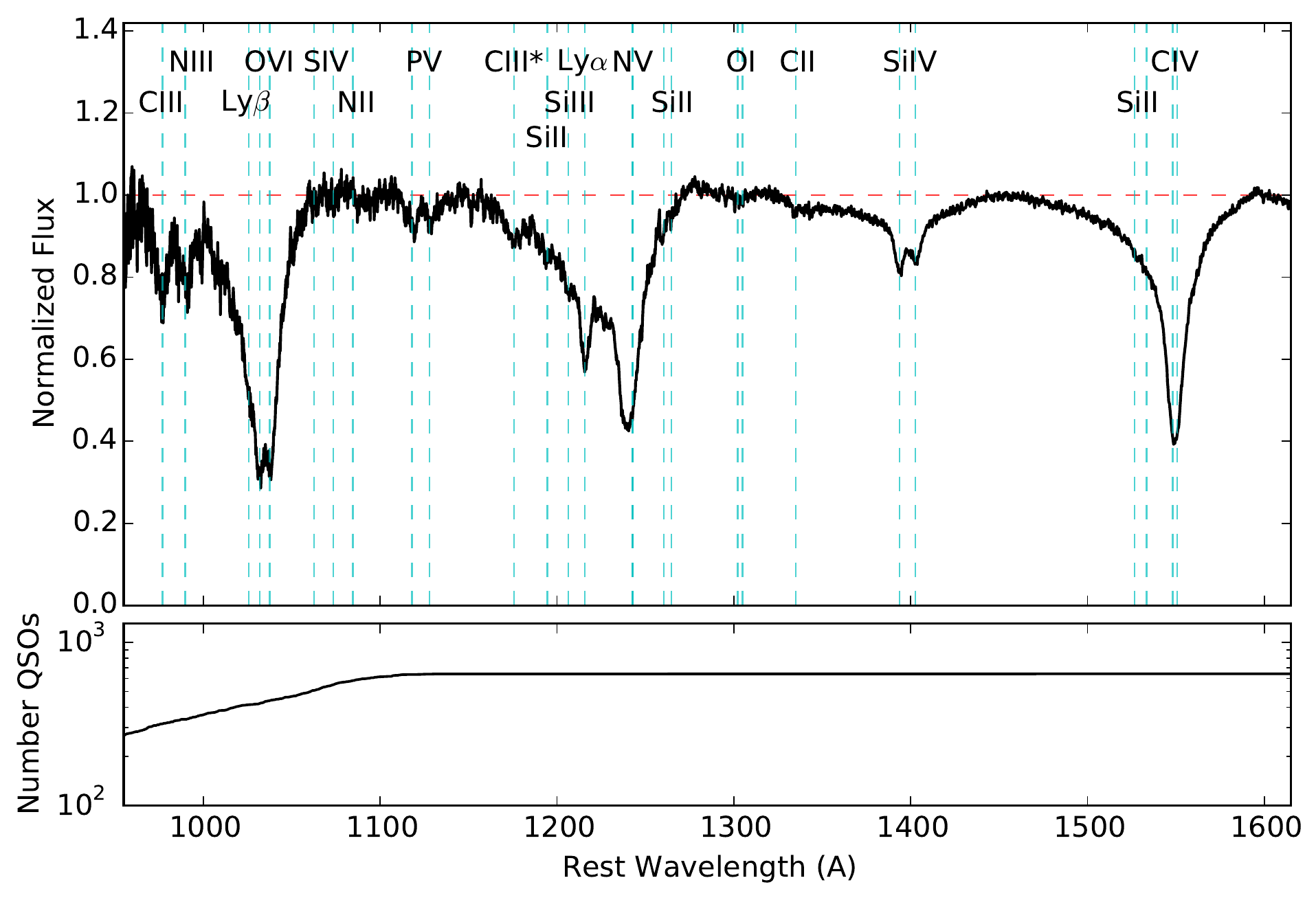}
\vskip -5pt
  \caption{{\it Top panel:} Normalized composite spectrum in the absorber frame for 641 BOSS quasars selected to have \civ\ BALs/mini-BALs with moderate strengths and minimum trough velocities $>$ 6000 \kms\ (Section 3.4). The dashed blue vertical lines with labels across the top mark absorption lines that are or might be present. The dashed red horizontal line marks the unabsorbed continuum. {\it Bottom panel:} Number of quasars contributing to the composite at each wavelength.}
    \label{fig:example_figure}
\end{figure}

The main result from Figure 5 is that the \ovi\ \lam 1034, \nv\ \lam 1240, and \civ\ \lam 1549 outflow lines have roughly similar strengths while \lya\ (in the blue wing of \nv ) is substantially weaker. This represents the typical situation for luminous quasars with BALs/mini-BALs crudely similar to the BAL in PDS~456. Another interesting result in Figure 5 is the significant presence of \pv\ \lam 1121 absorption and the resolved doublets with $\sim$1:1 ratios in \ovi , \pv , and \siiv . This supports the claim by Herbst et al. (in prep.) that BAL/mini-BAL outflows often have large total column densities and saturated absorption in all of the prominent lines. 

\subsection{Problems with the \lya\ Identification}

Figure 6 shows theoretical line-center optical depths from our Cloudy simulations (Section 3.2) for lines that should accompany a \lya\ BAL at 1346 \AA\ in PDS~456. In particular, the solid curves show the optical depths for lines with $b=5135$ \kms\ formed in clouds with a range of ionisation parameters, $U$, but with neutral hydrogen column densities held fixed at the derived value of $\log N_{\rm HI} ({\rm cm}^{-2}) = 15.35$ (Section 3.1). These predictions should be considered lower limits because $N_{\rm HI}$ derived from the data is a lower limit. Holding $N_{\rm HI}$ constant in the calculations leads to model clouds that have larger total hydrogen column densities, $N_{\rm H}$, at larger $U$. The $\log N_{\rm H}$ values for each $\log U$ are shown across the top of the figure. 

\begin{figure}
\begin{center}
	\includegraphics[width=\columnwidth]{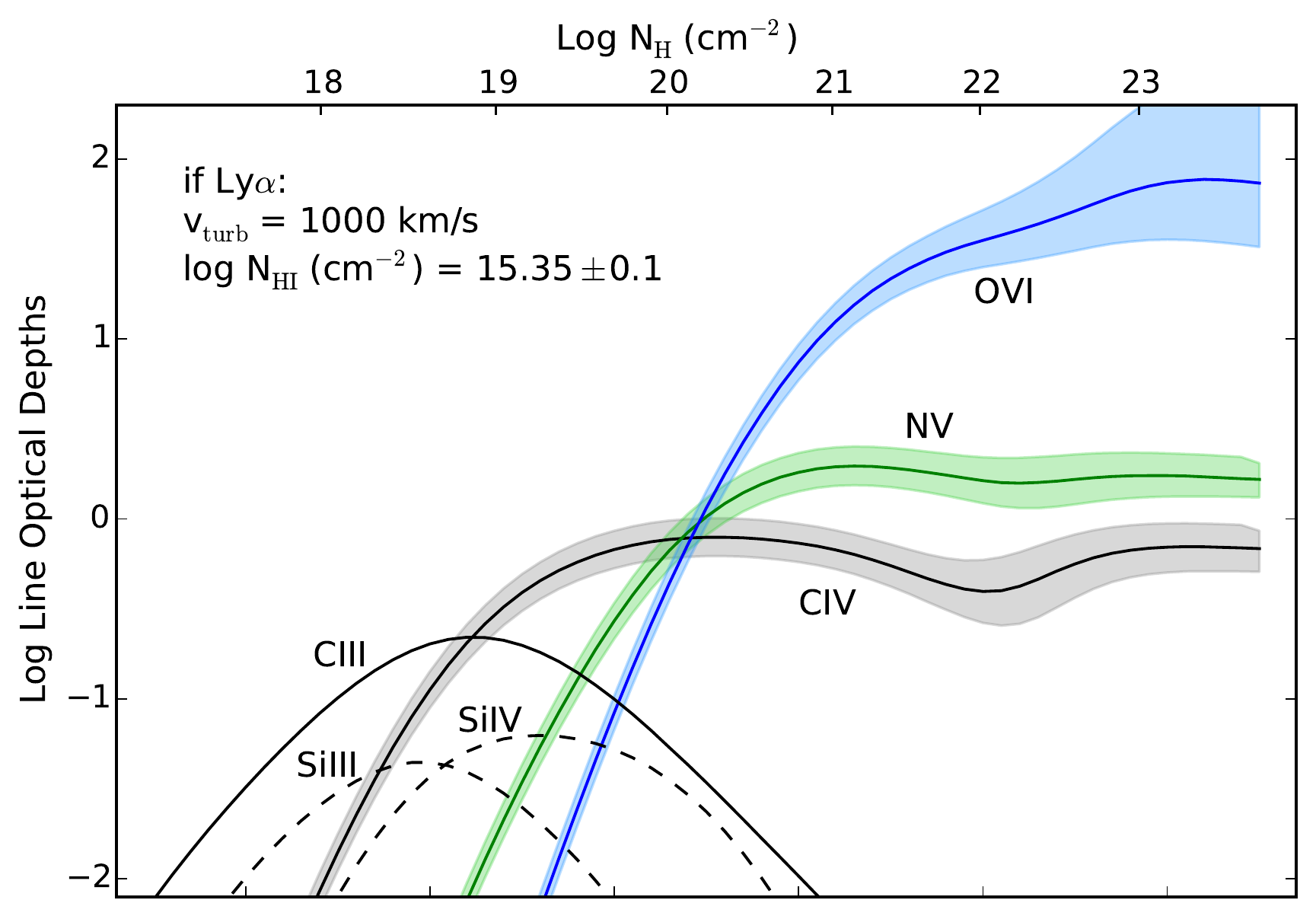}
	\includegraphics[width=\columnwidth]{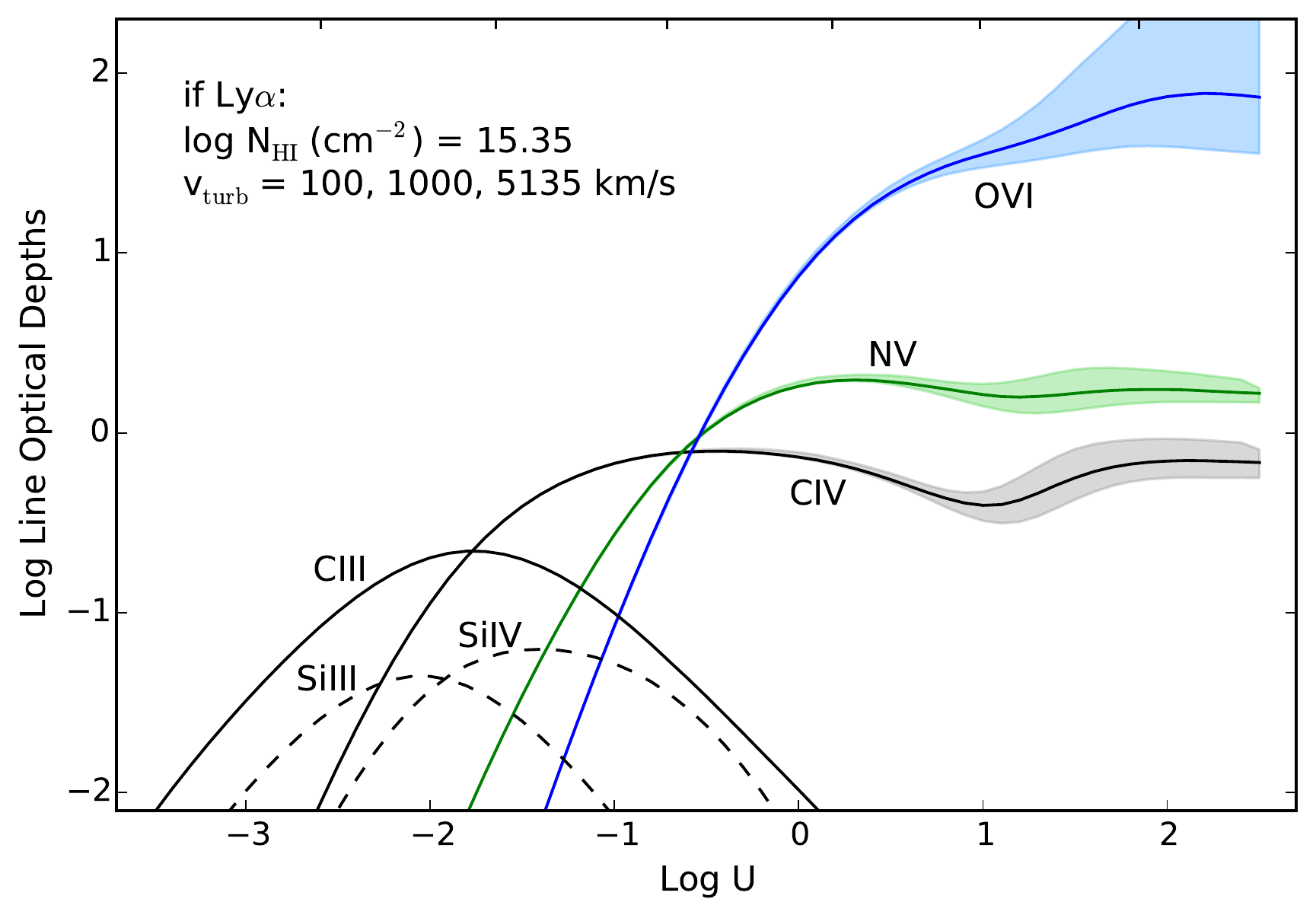}
\end{center}
\vskip -8pt
    \caption{Line-center optical depths versus ionisation parameter, $\log U$ (bottom axis), and total hydrogen column density, $\log N_{\rm H}$ (top axis) predicted by Cloudy simulations to accompany a \lya\ BAL like the observed feature in PDS 456 (with $\tau_o = 0.35$ and $b=5135$ \kms , Section 3.1). The dashed and dark solid curves are the same in both panels, representing model clouds with $\log N_{\rm HI} ({\rm cm}^{-2}) = 15.35$ and v$_{\rm turb} = 1000$ \kms . Shaded regions around the \civ , \nv, and \ovi\ curves show the ranges of optical depths that result from varying $\log N_{\rm HI}$ by $\pm$0.1 (top panel) or changing the turbulence from v$_{\rm turb}$ from 100 \kms\ to its maximum value, 5135 \kms , set by the observed BAL width (bottom panel). These predictions show that the \lya\ BAL identification would require $\log U < -2$ and uniquely low column densities, $\log N_{\rm H} ({\rm cm}^{-2}) < 18.7$, for a BAL outflow (see Section 3.5). }
    \label{fig:example_figure}
\end{figure}

The shaded regions around the curves for \civ , \nv , and \ovi\ in Figure 6 illustrate their dependence on the specific values of $N_{\rm HI}$ and the turbulent velocity, v$_{\rm turb}$, used in the calculations. The solid curves for these lines (and the solid and dashed curves for \ciii\ \lam 977, \siiii\ \lam 1206, and \siiv \lam 1398) correspond to fiducial parameters $\log N_{\rm HI} ({\rm cm}^{-2}) = 15.35$ and v$_{\rm turb} = 1000$ \kms . In the top panel, the shaded regions show the range of optical depths corresponding to the maximum uncertainty of $\pm 0.1$ dex in our $\log N_{\rm HI}$ measurement (see Section 3.1, where the smaller/larger $N_{\rm HI}$ value sets the lower/upper envelope to the shaded curves). The optical depth changes in response to these changes in $N_{\rm HI}$ are non-linear at large $U$ (and large $N_H$) because of radiative shielding in the far-UV that affects each ion differently. These shielding effects maintain significant optical depths in the moderate ions \civ , \nv , and \ovi\ at a fixed $N_{\rm HI}$ and  large $U$. In other calculations (not shown), we determine that the threshold for important shielding effects at large $U$ is roughly $\log N_{\rm HI} ({\rm cm}^{-2}) \gtrsim 14.8$. This is 0.55 dex lower than our lower limit to the \hi\ column density inferred from the observed BAL (Section 3.1), and therefore the predictions in Figure 6 that result from shielding at large $U$ should be applicable. 

Another parameter that can affect radiative shielding is the velocity dispersion inside the cloud, which we characterize by the turbulence velocity v$_{\rm turb}$. Large internal velocity dispersions can enhance the shielding by blending together numerous absorption lines in the far-UV, which in turn affects the ionisation structure deep in the cloud. The shaded regions around the \civ , \nv , and \ovi\ curves in the bottom panel of Figure 6 depict a range of results corresponding to v$_{\rm turb} = 100$ \kms , which yields negligible line shielding (the bottom envelope of the shaded regions), up to v$_{\rm turb} = 5135$ \kms , which yields the maximum line shielding allowed by the observed BAL width (the upper envelope of the shaded regions). We adopt an intermediate value v$_{\rm turb} = 1000$ \kms\ as our fiducial case (here and in all of our calculations below) because it yields conservatively small amounts of line shielding and, therefore, conservatively small predicted line optical depths in \civ , \nv , and \ovi\ at large $U$. 

The main result from Figure 6 is that a \lya\ BAL should be accompanied by significant broad \civ\ and \ovi\ absorption lines over a wide range of normal BAL outflow physical conditions. It is important to keep in mind that the predicted strengths of the \civ\ and \ovi\ lines are lower limits based on a lower limit on $N_{\rm HI}$ that follows from the assumption of no partial covering. Smaller values of $N_{\rm HI}$ are not allowed because they cannot produce the observed BAL attributed to \lya . Larger $N_H$ requiring larger $N_H$ (at a given $U$) would produce stronger \civ\ and \ovi\ absorption. The absence of these lines in the PDS~456 spectrum (Figures 2 and 3) means that the only conditions consistent with the \lya\ BAL identification are low degrees of ionisation, $\log U < -2$, and very low total column densities, $\log N_{\rm H} ({\rm cm}^{-2}) < 18.7$ that are $\gtrsim$3.5 orders of magnitude lower than expected for BAL outflows. For comparison, analysis of composite BAL quasar spectra by Herbst et al. (in prep.) suggests that {\it typical} BAL outflows have $\log N_{\rm H} ({\rm cm}^{-2}) \gtrsim 22.2$ (see also Sections 1 and 3.4). 

Another problem for the \lya\ identification is that a \lya -only BAL system would be, to our knowledge, unprecedented among observed BAL outflows. Figure 3 shows that the \civ\ and \ovi\ BALs expected to accompany \lya\ in PDS~456 would need to be $<$33 percent the strength of \lya\ to avoid detection. This is very different from the typical situation illustrated by Figure 5 and by other BAL composite spectra \citep[][Herbst et al., in prep.]{Baskin13} where the \civ\ and \ovi\ lines that are $\sim$2 to $\sim$3 times {\it stronger} than \lya . BAL surveys based on \civ\ BAL detections \citep[e.g.,][]{Trump06, Gibson09, Paris17} are naturally biased toward stronger \civ\ lines and they cannot detect \lya -only BAL systems if they exist. However, our visual inspections of the 641 outflow quasar spectra in our BOSS sample (Section 3.4) do not find any instances of \civ\ BALs weaker than, or even comparable to, the corresponding \lya\ absorption. 

A third argument against the \lya\ BAL identification stems from the weakening of the observed BAL between the HST observations in 2000 and 2014 (Figure 2). Variability in BALs and other outflow lines can be attributed to outflow clouds moving across our lines of sight or to changes in the ionisation caused by changes in the incident ionising flux \citep[cf.,][and refs. therein]{FilizAk13,Capellupo13,Capellupo14, Misawa14, Grier15, Arav12, Arav15, Rogerson16, Moravec17}. The evidence for BAL saturation discussed in Section 3.1 and Section 3.4 might favor crossing clouds in many situations. However, if the BAL changes in PDS~456 were caused by ionisation changes and the ionisation was initially low to produce \lya\ absorption without accompanying \civ\ and \ovi\ lines, then the only way to make \lya\ weaker at a fixed $N_H$ is with larger ionisation parameters that should produce stronger \civ\ and \ovi\ absorption. The continued absence of these accompanying lines in the 2014 (Figure 2) thus makes the \lya\ identification more difficult because the values of $N_{\rm H}$ and $U$ in 2000 would need to be several times lower than the already-low upper limits deduced from Figure 6. 

\subsection{Problems with the \nv\ \lam 1240 Identification}

The top panel in Figure 7 shows Cloudy predictions for the line-center optical depths that should accompany the observed BAL if it is \nv\ \lam 1240. As in Figure 6, the calculations assume a cloud velocity dispersion v$_{\rm turb} = 1000$ \kms\ (to include moderate amounts of line shielding) but line optical depths derived for $b=5135$ \kms . The total column densities in the model clouds are adjusted to yield an \nv\ column density, $\log N(\nv  ) ({\rm cm}^{-2}) = 15.62$, fixed to the value inferred from the observed BAL (Section 3.1). The resulting optical depths are again only lower limits because $N(\nv )$ is a lower limit. 

\begin{figure}
\begin{flushright}
	\includegraphics[width=\columnwidth]{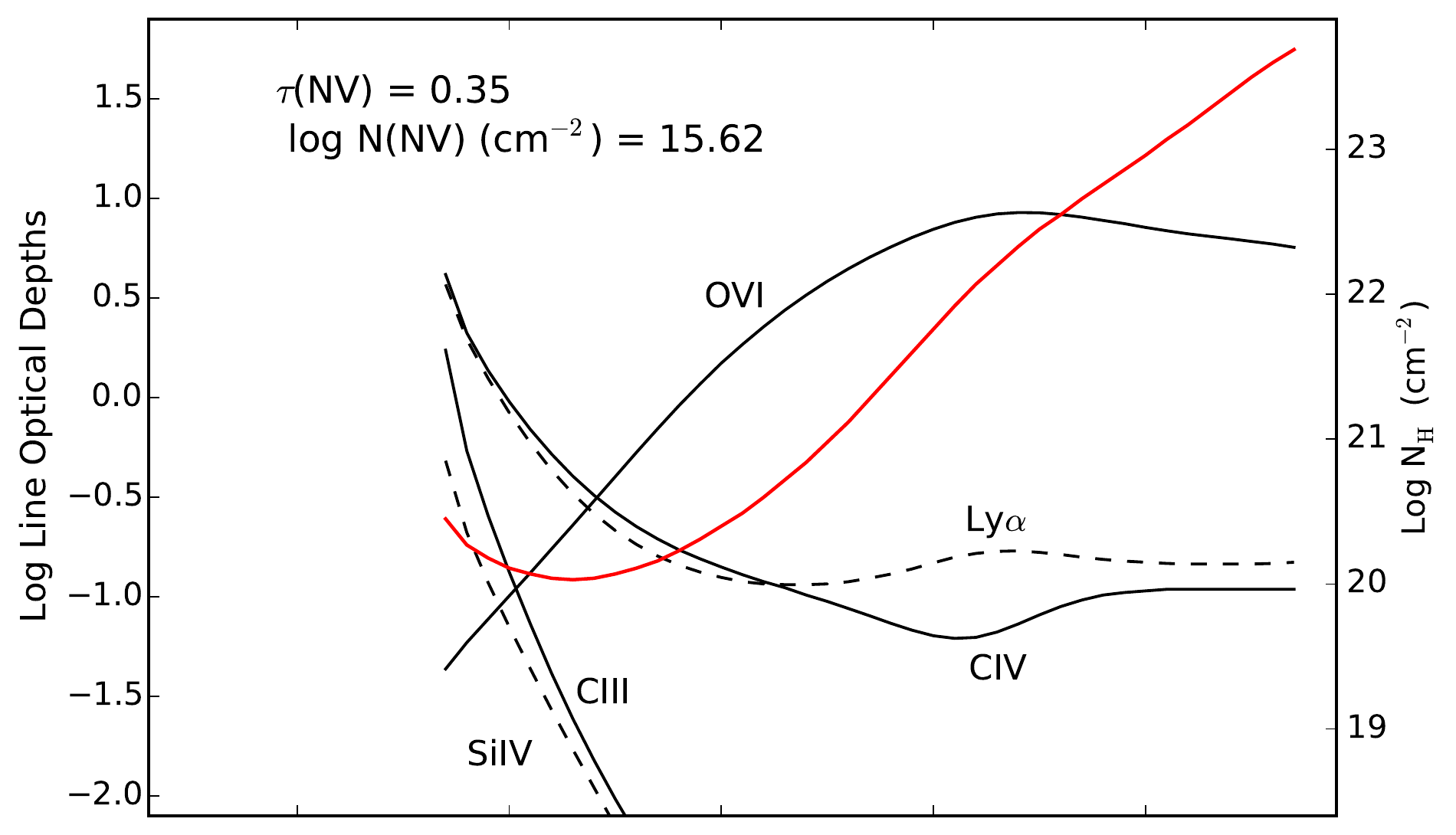}
	\includegraphics[scale=0.437]{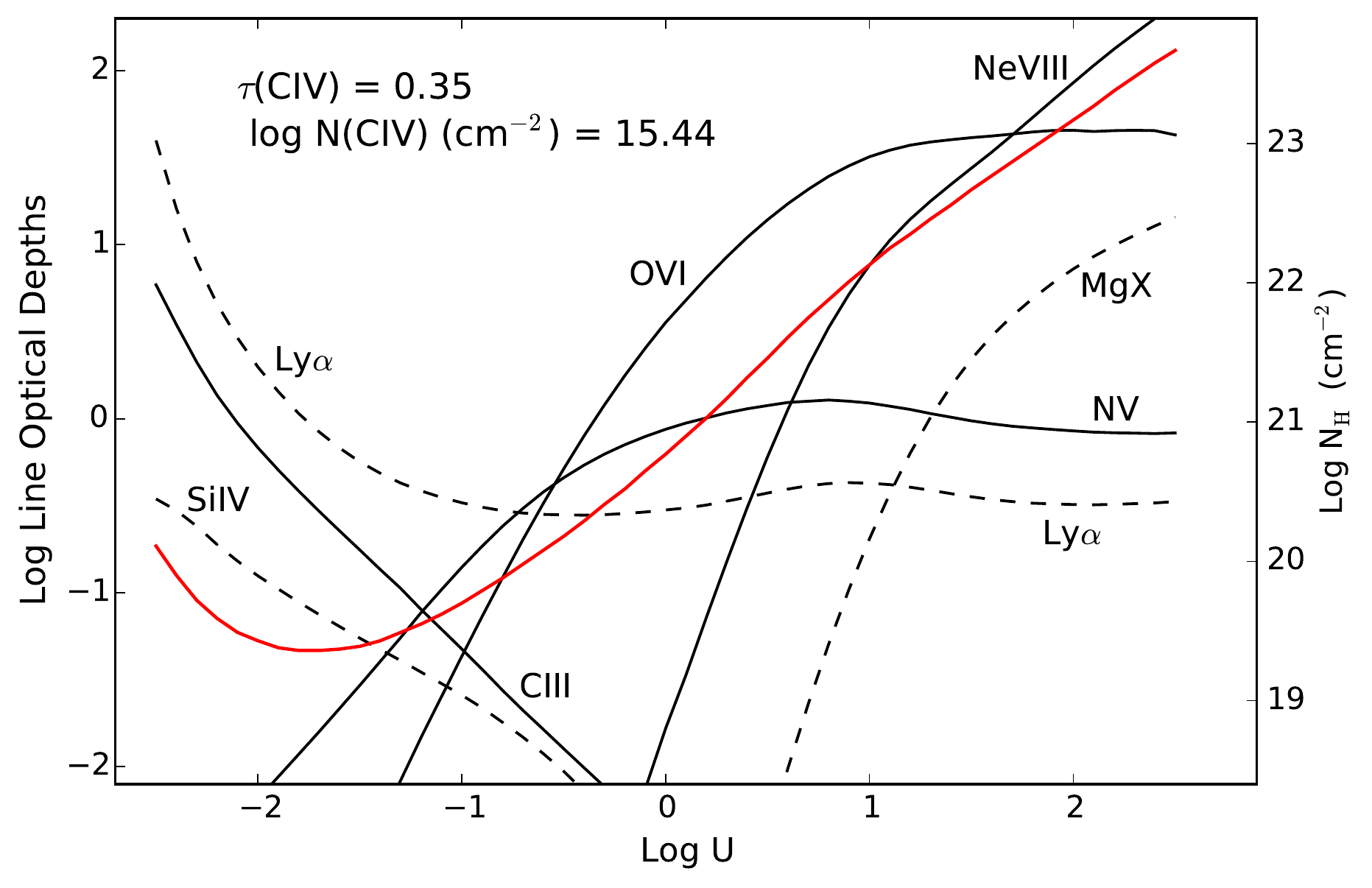}
\end{flushright}
\vskip -7pt
    \caption{Line-center optical depths (black curves) and total hydrogen column densities, $\log N_{\rm H}$ (red curve), versus ionisation parameter, $\log U$, predicted by Cloudy to accompany the observed BAL (with $\tau_o = 0.35$ and $b=5135$ \kms ; Section 3.1) if it is attributed to \nv\ (top panel) or \civ\ (bottom panel). The ionic column densities held fixed in the model clouds are shown in the upper left of each panel. These predictions and existing spectra (Figures 1--3) rule out low degrees of ionisation for the \nv\ BAL system, and they can guide future observations to test both BAL identifications further (see Section 3.6 and Section 3.7).}
    \label{fig:example_figure}
\end{figure}

The observed BAL at 1346 \AA\ is compatible with the \nv\ identification only if the accompanying \civ\ line is undetectable. This requires a ratio of \nv / \civ\ absorption line strengths $\gtrsim 3$ (Section 3.1, Figure 3). Figure 7 shows that moderate-to-low outflow ionisations with $\log U \lesssim 0$, corresponding to total columns $\log N_{\rm H} ({\rm cm}^{-2}) \lesssim  20.2$, are firmly ruled out for the \nv\ BAL identification by the predictions for strong \civ\ absorption. At higher ionisations, the predicted optical depth ratio hovers around $\tau (\nv )/\tau (\civ) \sim 3$--5 (Figure 7), which is  marginally consistent with the observed spectrum. However, any significant partial covering effects would push the observed \nv /\civ\ line depth ratio toward unity, such that \civ\ should be detectable and the \nv\ BAL identification is ruled out. 

Figures 4 and 5 show more directly that absorption line ratios \nv /\civ\ $\gtrsim 3$ needed for the \nv\ BAL identification are, at best, extremely rare in observed quasar outflows. In particular, the composite spectrum (Figure 5) shows that typical weak-to-moderate BAL/mini-BAL systems have a median line depth ratio of \nv /\civ$\;\sim 1\pm 0.1$. Individual outflows  can have larger \nv /\civ\ ratios. Measurements in the literature of BALs/mini-BALs at speeds up to nearly 0.2$c$ reveal \nv /\civ\ depth ratios in the range $\sim$0.5 to 1.5 \citep{Steidel90, Jannuzi96, Hamann97e, Hamann97b, Barlow97, Telfer98, Sabra03, Paola08, Paola11, Paola13, Rogerson16, Moravec17}. Our visual inspections of the 641 outflow quasars in Section 3.4 find crudely $\sim$2 percent of cases with \nv /\civ\ $\gtrsim 2$. Figure 4 shows the most extreme well-measured examples. It is difficult to assess the \nv /\civ\  line ratios quantitatively in these spectra due to blending problems in the \lya\ forest. However, the most extreme cases\footnote{We need to acknowledge here that hypothetical quasars with strong \nv\ BALs but negligible \civ\ absorption cannot appear in our BOSS sample because it relies on \civ\ line detections (via AI and BI).} shown in Figure 4 also appear to be inconsistent with the \nv\ BAL identification in PDS~456. 

Another constraint on the \nv\ BAL identification is that it should be accompanied by strong \ovi\ absorption at observed wavelength $\sim$1122 \AA . This is based on both theoretical predictions (Figure 7) and observations of other outflow quasars (Section 3.4 and references immediately above). The wavelength 1122 \AA\ is just outside the wavelength coverage of the 2000 STIS spectrum; however, there is no evidence for a broad \ovi\ line wing that could be measurable if this line was present (Figure 3). The 2014 COS spectrum does cover these \ovi\ wavelengths and clearly shows no signs of absorption there (Figure 2), but the BAL at 1346 \AA\ was also much weaker in this spectrum. Thus the constraints provided by absence of \ovi\ in existing spectra are weak, but they also do not support the \nv\ BAL identification.

\subsection{The Case for \civ\ \lam 1549 at 0.30c}

One argument favoring the \civ\ BAL identification in PDS~456 is that it is readily compatible with existing spectra (Figures 1--3).  It has none of the problems described above for \lya\ and \nv\ because no other lines are expected within the wavelength coverage. Observations of high-velocity BALs/mini-BALs in other quasars indicate that the outflow ionisations are generally high and that the most prominent lines accompanying \civ\ should be \nv\ and \ovi\  (Section 3.4, Section 3.6, \citealt{Hamann13}). The bottom panel in Figure 7 shows specific theoretical predictions for the optical depths in these and other lines that should accompany the observed BAL if it is \civ . The most observationally-accessible lines in this plot are \lya\ at low ionisations and \nv\ at high ionisations, both at predicted observer-frame wavelengths near $\sim$1068 \AA . Future observations might test for this absorption, but that will be difficult due to the short wavelength and severe reddening (Section 2). The situation is worse for other corroborating lines. In particular, the corresponding \siiv\ \lam 1398 line would be at $\sim$1216 \AA\ observed, which is directly on top of the Galactic damped \lya\ line, and the \ovi\ and \neviii\ \lam 774 lines, which should be strong at high ionisations are at inaccessible observed wavelengths ($\sim$899 \AA\ and $\sim$673 \AA , respectively) due to Galactic Lyman limit absorption. 

A second circumstantial argument favoring the \civ\ identification is that its velocity v$\;\approx 0.30c$ is similar to the speeds measured for X-ray outflow of PDS~456. The X-ray outflow is highly variable with multiple velocity components, but the main component identified by Fe K-shell absorption has measured velocities in the range v$\;\sim 0.25$--0.34$c$ with a typical value near $\sim$0.3$c$ \citep{Reeves14, Reeves16, Nardini15, Matzeu17}. The \civ\ BAL might provide evidence for a physical relationship between the UV and X-ray outflows in PDS~456 (see Section 4 below). 

A third argument is that relativistic \civ\ BALs/mini-BALs at speeds approaching v$\;\sim 0.2c$ have already been measured in a growing number of luminous quasars \citep[][Rodriguez Hidalgo et al., in prep.]{Jannuzi96, Hamann97, Hamann13, Paola08, Paola11, Rogerson16}. The \civ\ BAL at v$\;\approx 0.30c$ in PDS~456 would set a new speed record for UV outflows, but it is not so dramatic to be a paradigm shift for our understanding of these outflows. The width and depth of the BAL in PDS~456 is roughly similar to these other high-velocity outflow features. We also note that our tentative detection of another weak \civ\ BAL in PDS~456 at v$\;\sim 0.19c$ (Figure 2, Section 3.2) provides additional, albeit tentative, evidence that a relativistic UV outflow is present in PDS~456.

\section{Summary \& Discussion}

The UV spectrum of PDS~456 obtained with HST-STIS in 2000 has a distinct BAL at $\sim$1346 \AA\ that we identify as \civ\ \lam 1549 at velocity shift v$\; \approx 0.30c$, FWHM~$\approx 8550$ \kms , and minimum optical depth $\tau_o = 0.35$ (Section 3.1). The \civ\ identification rests on its compatibility with existing spectra and the absence of lines that should accompany the alternative identifications, \lya\ \lam 1216 and \nv\ \lam 1240. \lya\ is compatible with the observed BAL only if the outflow has a low degree of ionisation, $\log U < -2$, and a very low total column density, $\log N_{\rm H} ({\rm cm}^{-2}) <18.7$, that would be unprecedented in BAL outflow studies, e.g., several orders of magnitude below recent estimates (Section 3.5). The \nv\ identification might be consistent with absence of accompanying \civ\ absorption if the gas is highly ionised and it pushes the boundary of observed \nv /\civ\ line strengths beyond what we find in our comparison of 641 outflow quasars in BOSS (Section 3.6). However, this situation should produce strong \ovi\ absorption, which is not well constrained in existing spectra of PDS~456 but it appears to be absent (Figure 3). Thus the \nv\ identification is also strongly disfavored. 

The \civ\ BAL identification has none of these problems (Section 3.7). It would mark the fastest UV outflow line ever reported, but its velocity is consistent with the X-ray outflow in PDS~456 (see Section 1 and below) and not dramatically different from the relativistic \civ\ BALs/mini-BALs already measured at speeds approaching $\sim$0.2$c$ in other quasars \citep[][Rodriguez Hidalgo et al., in prep.]{Jannuzi96, Hamann97, Hamann13, Paola08, Paola11, Rogerson16}. The \civ\ identification is also weakly supported by our tentative identification of an additional \civ\ BAL feature at v$\,\sim 0.19c$ (, Figure 2, Section 3.2). Broad UV outflow lines at speeds near $0.3c$ are surely very rare based on the rarity of such lines at $\sim$0.1--0.2$c$ (see refs. above). However, the incidence of \civ\ BALs/mini-BALs at v$\;> 0.2c$ is not known because they have not been searched for in large quasar surveys like SDSS/BOSS and, in any case, they could easily be missed due to blends with unrelated lines in the \lya\ forest \citep[e.g., see the search for \pv\ \lam 1121 BALs by][]{Capellupo17}. 

The location of the \civ\ BAL outflow in PDS~456 is a critical unknown. The range of ionizations consistent with \civ\ absorption might favor lower ionizations and larger distances from the black hole than the X-ray outflow, perhaps at radii of order $\sim$1 pc as inferred from some variability studies of \civ\ BALs/mini-BALs in other quasars \citep[][McGraw et al., submitted]{Hamann13, Capellupo14, Moravec17}. However, larger distances do not necessarily produce lower ionisations, e.g., if acceleration causes the outflow densities to drop faster than the $1/r^2$ behaviour expected from free expansion at a constant speed. Moreover, the relativistic speed of the \civ\ BAL at v$\; \sim 0.30c$ indicates that the UV outflow originated with the X-ray UFO very close to the black hole. If the measured flow speeds are similar to the gravitational escape speed at the launch radius, then the launch point is roughly at $r\sim 20$--$30\, r_g$ \citep{Nardini15, Matzeu17}. Also note that a highly-ionised X-ray outflow launched from this radius and expanding freely outward into a fixed solid angle will not necessarily become less ionised at larger distances (for \civ\ absorption) because the $1/r^2$ dilution of the radiation field is balanced by a $1/r^2$ decline in the densities to yield a constant ionisation parameter. Lower ionisations will occur if the inner regions of the outflow radiatively shield the material downstream and/or if there are clumps with enhanced densities relative to the ambient flow. Clumping and shielding can occur to the same effect at almost any radius, and there is already evidence for dense clumps in the X-ray outflow of PDS~456 (based on lower ionizations and partial covering in the soft X-ray absorber, \citealt{Reeves16,Matzeu16}; see also \citealt{Gofford14, Nardini15, Hagino15}).

It is therefore an intriguing possibility that the \civ\ BAL forms directly within, or in close proximity to, the relativistic X-ray outflow. 
Our Cloudy simulations (Figure 7) show that the \civ\ BAL in PDS~456 could form over a wide range of physical conditions, including very high ionisations where \civ\ is just a trace constituent. This situation is illustrated in Figure 8, which plots the ionisation structure in a single model cloud with $\log U = 1.7$. This cloud reaches the observed minimum BAL optical depth $\tau_o (\civ ) = 0.35$, along with $\tau_o (\nv ) = 0.90$ and $\tau_o (\ovi ) = 43$, at total column density $\log N_{\rm H}({\rm cm}^{-2}) = 22.9$ (represented by the unshaded left-hand portion of Figure 8). This front portion of the cloud matches our calculations in the bottom panel of Figure 7. In this environment, the \civ\ and \nv\ ion fractions are everywhere $\lesssim$$4\times 10^{-4}$ and $\lesssim$$3\times 10^{-3}$, respectively, and the dominant form of oxygen is \ovii . This highly-ionsed environment capable of \civ\ \lam 1549 absorption will also produce saturated absorption in higher ion lines such as \neviii\ \lam 774 and \mgx\ \lam 615 (see also $\log U = 1.7$ Figure 7) and at the bound-free edges of \ovii\ and \oviii\ in soft X-rays. 

\begin{figure}
	\includegraphics[width=\columnwidth]{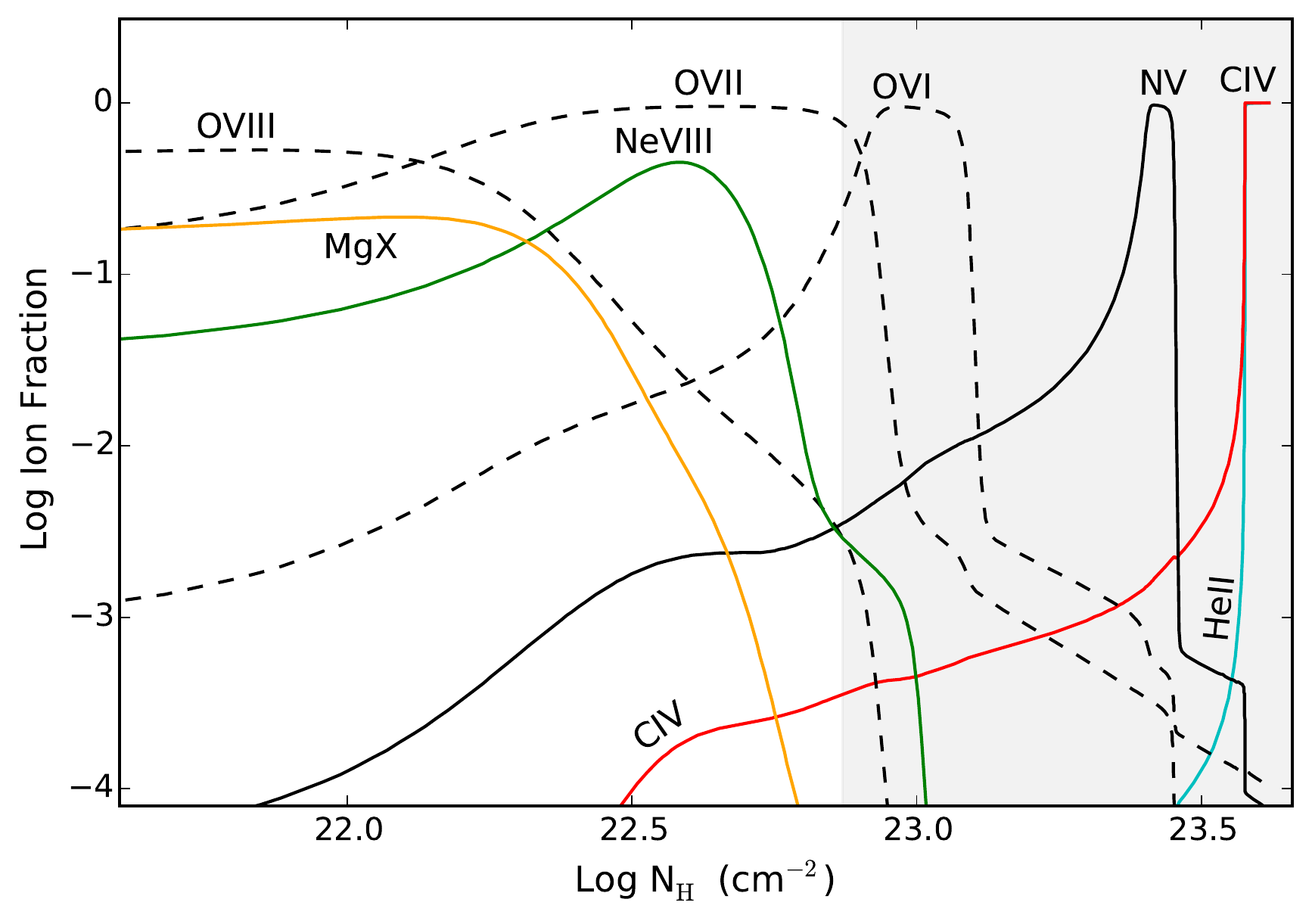}
\vskip -7pt
    \caption{Ionisation fractions versus total hydrogen column density, $\log N_{\rm H}$, in a single model cloud with ionisation parameter, $\log U = 1.7$. This cloud produces the minimum observed BAL optical depth in \civ\, $\tau_o (\civ ) = 0.35$, at $\log N_{\rm H}({\rm cm}^{-2}) \approx 22.9$, such that the unshaded left-hand region represents the model cloud with $\log U = 1.7$ in Figure 7. This environment (at $\log N_{\rm H}({\rm cm}^{-2}) \leq 22.9$) is highly ionised and optically thick at the bound-free edges of \ovii , \oviii , and \neviii , with only trace amounts of the lower ions \civ , \nv , and \ovi . The lower ions dominate at larger column densities (grey shaded region) where there is more radiative shielding, but these lower ionisations are not needed for the \civ\ BAL in PDS~456.}
    \label{fig:example_figure}
\end{figure}

Detailed comparisons between the UV and X-ray outflows in PDS~456 are beyond the scope of the present study. They are subject to uncertainties caused by the outflow variabilities and by unknowns in the shape of the ionising spectrum and the spatial locations of different outflow components relative to the ionizing far-UV and X-ray emission sources. Here we note simply that recent estimates of the X-ray outflow ionisation parameters\footnote{The ionisation parameter quoted in the X-ray studies is $\xi = L_{\rm ion}/n_Hr^2$ from \cite{Tarter69}, where $L_{\rm ion}$ is the quasar luminosity from 1 to 1000 Rydberg. For $\xi$ in units of ergs cm s$^{-1}$ and the continuum shape used in our Cloudy simulations (Section 3.3), the conversion is $\log U \approx \log \xi - 1.2$.} are roughly in the range $\log U \sim 4.3$ to 4.8 for the K-shell outflow \citep{Gofford14, Nardini15} and $\log U \sim 2.8$ for the soft X-ray absorber \citep{Reeves16}. The ionisation parameters needed for the \civ\ BAL, with upper limit $\log U \lesssim 2.4$ (Figure 7), might occur in dense clumps embedded in the X-ray outflow if the density enhancements are $\gtrsim$0.4 dex relative to the soft X-ray absorber or $\gtrsim$2 dex relative to the Fe K-shell outflow. 

The ionization structure of the UV outflow might resemble the quasar SBS~1542+541 (redshift $z\sim 2.36$), which has measured UV BALs ranging in ionization from \civ\ up to \mgx\ and \sixii\ \lam 510 \citep{Telfer98}. The lines in this quasar appear saturated with shallow troughs that reveal ion-dependent line-of-sight covering fractions from $\sim$15 percent in \civ\ to $\sim$50 percent in the higher ions. \cite{Telfer98} infer from this a 2-zone outflow that could span a decade or more in ionisation parameter, with the \civ\ lines forming in small clumps embedded in a more highly-ionised outflow medium. \cite{Telfer98} also note that this outflow should produce substantial bound-free absorption by \ovii\ and \oviii\ in soft X-rays. 

The general picture of clumpy outflows has become commonplace in quasar outflow studies. High-quality observations of UV outflow lines often provide evidence for clumpy multi-phase outflow structures with a range of covering fractions \citep{Moravec17, Misawa14b, Hamann11, Hamann04, Gabel05, Leighly15, Leighly11, Leighly09, Misawa07c, Arav08, Arav05, deKool02, Ganguly99}. Detailed studies of some bright Seyfert 1 galaxies \citep[e.g.,][]{Kaspi02, Netzer03, Gabel05b} clearly demonstrate that the UV and X-ray absorption features can form together in complex outflows, with indications that the lower-ionisation UV lines identify clumps or filaments embedded in the X-ray outflow. Clumpy outflow structures are also predicted by recent numerical simulations \citep{Sim10, Takeuchi13, Waters17} and by considerations of the radiative forces that can compress the outflows into small substructures \citep{Stern14b}. There are also theoretical arguments requiring dense clumps to moderate the outflow ionizations in the absence of significant radiative shielding \citep{deKool97, Hamann13}. 

If the \civ -absorbing gas in PDS~456 is indeed embedded in the X-ray outflow, it could have important implications for the outflow energetics. It is a well-known problem for X-ray UFOs \citep[e.g.,][]{Tombesi11, Tombesi13, Gofford15} that their high ionizations lead to low opacities and inefficient radiative acceleration \citep{Gofford13, Gofford14, Higginbottom14}. Harnessing the full radiative power of the quasar to drive these outflows might require opacities beyond electron scattering at UV/far-UV wavelengths, near the peak of the quasar spectral energy distributions. \cite{Matzeu17} showed recently that the outflow speeds in PDS~456 correlate with the variable X-ray luminosity, consistent with radiative acceleration \citep[see also][]{Saez11}. If radiative forces are important, then dense clumps with lower ionisations embedded in the X-ray outflows might be important to boost the opacities for radiative driving \citep{Laor14, Hagino15}. The \civ\ BAL at v$\;\approx 0.30c$ in PDS~456 could be the first direct observational evidence for this idea. 

PDS~456 is a remarkable object with powerful accretion-disk outflows launched from a range of at least two decades in disk radii, from $\sim$0.001 pc for the X-ray outflow to $\sim$0.3 pc for the blueshifted CIV broad emission line \citep[Section 1,][]{OBrien05, Gofford14}. More work is needed to understand how the UV BAL fits into this outflow environment and, specifically, to test the \civ\ BAL identification. A search for corroborating \nv\ absorption at $\sim$1077 \AA\ will be difficult due to Galactic reddening (Section 2) and the poor sensitivity of current instruments at these wavelengths (Section 3.7). However, the alternative BAL identifications, \nv\ and \lya , might be ruled out or confirmed more easily by searching for \ovi\ absorption near 1122 \AA\ (Section 3.6, although the BAL at 1346 \AA\ needs to be present for a meaningful test). Our team has ongoing programs to extend the UV wavelength coverage and monitor PDS~456 in the UV and X-rays that will be described in future papers. 

\section*{Acknowledgements}

We are grateful to Gary Ferland and the Cloudy development team for their continued support and public dispersement of the spectral synthesis code Cloudy. FH also thanks Nahum Arav, Chris Done, Gerald Kriss, and Paola Rodriguez Hidalgo for helpful discussions. We are grateful to the referee, Paul Hewett, for helpful comments that improved this manuscript. JNR acknowledges the support via HST grant HST-GO-14477.001-A. EN acknowledges funding from the European Union's Horizon 2020 research and innovation programme under the Marie Sklodowska-Curie grant agreement no. 664931.




\bibliographystyle{mnras}
\bibliography{../bibliography} 


\bsp	
\label{lastpage}
\end{document}